\documentclass[conference]{IEEEtran}
\IEEEoverridecommandlockouts 

\usepackage{cite}
\usepackage{graphicx}
\usepackage{textcomp}
\usepackage{xcolor}
\usepackage{array}
\usepackage{booktabs}
\usepackage{hyperref}
\usepackage[nolist]{acronym}
\usepackage{orcidlink}
\usepackage{balance}
\usepackage{xurl}
\usepackage{fancyhdr}
\def\BibTeX{{\rm B\kern-.05em{\sc i\kern-.025em b}\kern-.08em
    T\kern-.1667em\lower.7ex\hbox{E}\kern-.125emX}}

\begin{document}

\begin{acronym}
\acro{wui}[WUI]{Wildland-Urban Interface}
\acro{nasa}[NASA]{National Aeronautics and Space Administration}
\acro{usgs}[USGS]{United States Geological Survey}
\acro{noaa}[NOAA]{National Oceanic and Atmospheric Administration}
\acro{mtbs}[MTBS]{Monitoring Trends in Burn Severity}
\acro{crs}[CRS]{Coordinate Reference System}
\acro{feds}[FEDS]{Fire Event Data Suite}
\acro{viirs}[VIIRS]{Visible Infrared Imaging Radiometer Suite}
\acro{hrrr}[HRRR]{High-Resolution Rapid Refresh}
\acro{aws}[AWS]{Amazon Web Services}
\acro{gee}[GEE]{Google Earth Engine}
\acro{lai}[LAI]{Leaf Area Index}
\acro{frp}[FRP]{Fire Radiative Power}
\acro{firms}[FIRMS]{Fire Information for Resource Management System}
\acro{threedep}[3DEP]{3D Elevation Program}
\acro{landfire}[LANDFIRE]{Landscape Fire and Resource Management Planning Tools}
\acro{epsg}[EPSG]{European Petroleum Survey Group}
\acro{utm}[UTM]{Universal Transverse Mercator}
\acro{ai}[AI]{Artificial Intelligence}
\end{acronym}

\title{FireDataForge: A Unified Framework for Multi-Source Wildfire Data Retrieval and Integration}

\author{%
\IEEEauthorblockN{%
Zeyu Xia\textsuperscript{*}\,\orcidlink{0000-0003-0234-5857},\quad
Lexie Chen\textsuperscript{*}\,\orcidlink{0009-0004-6682-1474},\quad
Ye Liu\textsuperscript{\dag}\,\orcidlink{0000-0001-5131-8412},\quad
Huilin Huang\textsuperscript{*}\,\orcidlink{0000-0002-7328-6738}}
\IEEEauthorblockA{%
\textsuperscript{*}\textit{University of Virginia}, Charlottesville, VA, USA \\
\textsuperscript{\dag}\textit{Pacific Northwest National Laboratory}, Richland, WA, USA \\
zeyu.xia@virginia.edu, dkt4kr@virginia.edu, ye.liu@pnnl.gov, mgh4rc@virginia.edu}
\IEEEauthorblockA{\parbox{0.9\textwidth}{\centering\footnotesize\vspace{1ex}
\textbf{Author Accepted Manuscript.} Cite as: Z.~Xia, L.~Chen, Y.~Liu, and H.~Huang, ``FireDataForge: A Unified Framework for Multi-Source Wildfire Data Retrieval and Integration,'' in \textit{2026 IEEE International Conference on Information Reuse and Integration for Data Science (IRI)}, Seattle, WA, USA, 2026, pp.~528--531, doi: \href{https://doi.org/10.1109/IRI69576.2026.00103}{10.1109/IRI69576.2026.00103}.}}
\thanks{\copyright~2026 IEEE\@. Personal use of this material is permitted. Permission from IEEE must be obtained for all other uses, in any current or future media, including reprinting/republishing this material for advertising or promotional purposes, creating new collective works, for resale or redistribution to servers or lists, or reuse of any copyrighted component of this work in other works.}
}

\maketitle
\thispagestyle{fancy}

\begin{abstract}
Wildfire research, modeling, and education require geospatial data from multiple sources that vary in formats, coordinate systems, spatial resolutions, and temporal cadences. This preprocessing burden limits reproducible reuse. We present FireDataForge, an open-source Python framework that automates retrieval and harmonization of 11 wildfire-related sources spanning fire behavior, weather, land cover, vegetation, elevation, built environment, wildland-urban interface, fire history, and satellite imagery. Given an MTBS Event ID, FireDataForge retrieves relevant datasets, aligns them to a common grid, and outputs analysis-ready NumPy arrays with embedded metadata. Batch processing of historical fires demonstrates support for fire behavior simulation, educational visualization, machine learning, and AI-assisted wildfire analysis.
\end{abstract}


\section{Introduction}

Across the United States, increasingly warm and dry conditions have contributed to rising wildfire frequency, intensity, burned area, and rates of spread~\cite{burtonGlobalBurnedArea2024,parkAttributingHumanMortality2024}. Expanding development in fire-prone landscapes has also increased ignition risks and \ac{wui} exposure. Understanding and predicting wildfire behavior requires integrating geospatial data on fuel conditions, meteorology, topography, and human activity~\cite{scottWildfireRiskAssessment2013a,altintasActionableFireModeling2025}.

These datasets are fragmented across agencies (e.g., \ac{nasa}, \ac{usgs}, \ac{noaa}), stored in heterogeneous formats, and provided at different spatial and temporal resolutions.\footnote{NASA Earth Science Data Systems, ``Data Formats,'' Earthdata, \url{https://www.earthdata.nasa.gov/learn/earth-observation-data-basics/data-formats}.} Researchers must therefore manually retrieve, preprocess, and align data before analysis, limiting reproducibility, scalability, and accessibility.

We present \textbf{FireDataForge}, a Python framework that automates multi-source wildfire data retrieval, harmonization, and integration for research, education, and model development. Given a \ac{mtbs} Event ID~\cite{eidenshinkProjectMonitoringTrends2007}, FireDataForge retrieves relevant datasets, reprojects and resamples them to a common grid, aligns them temporally, and outputs metadata-rich NumPy arrays. These outputs support physics-based and data-driven fire-spread modeling, 2D/3D visualization of fire progression and \ac{wui} exposure, machine learning, and \acs{ai}-assisted historical-event analysis.

This work contributes (1) a unified pipeline integrating heterogeneous wildfire datasets, (2) automated spatial and temporal alignment with configurable resolution and coordinate systems, and (3) standardized, reusable outputs supporting these downstream uses.

\begin{figure*}[!t]
\centering
\includegraphics[width=.96\linewidth]{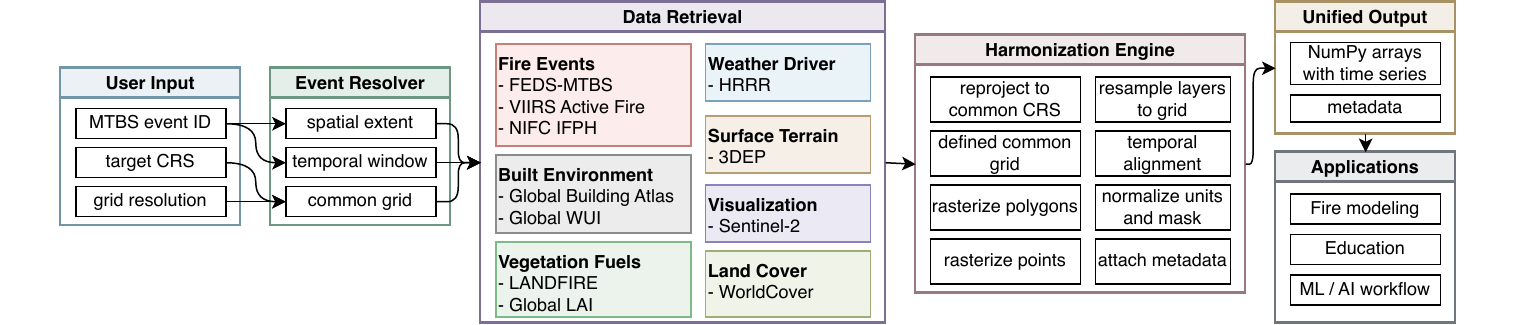}
\caption{FireDataForge workflow from an MTBS Event ID to harmonized, metadata-rich NumPy outputs.}\label{fig:pipeline}
\end{figure*}

\section{Related Work}

Cloud platforms and access libraries improve retrieval of geospatial inputs: \ac{gee}~\cite{gorelickGoogleEarthEngine2017} and geemap~\cite{wuGeemapPythonPackage2020} support satellite and derived products, while Herbie~\cite{blaylockHerbieRetrieveNumerical2026} retrieves numerical weather archives such as \ac{hrrr} on \ac{aws}. These tools remain source-oriented and require custom workflow scripts to produce event-centered, model-ready wildfire datasets.

Application systems include visualization platforms~\cite{altintasActionableFireModeling2025} and multimodal fire-risk fusion frameworks~\cite{yuanFireRiskMultiDynamicMultimodal2025a}. In parallel, deep learning models increasingly fuse remote sensing and environmental data for wildfire risk and spread prediction~\cite{xuDeepLearningWildfire2025,zhouComparativeInterpretativeAnalysis2025}. Curated ML datasets such as BCWildfire~\cite{xuBCWildfireLongtermMultifactor2026}, Next Day Wildfire Spread~\cite{huotNextDayWildfire2022}, and Mesogeos~\cite{kondylatosMesogeosMultipurposeDataset2023} align topography, vegetation, weather, and fire data for fixed domains, periods, grids, and cadences. FireDataForge is complementary: it generates analysis-ready data on demand from an \ac{mtbs} Event ID, maps sources to a user-specified \ac{crs} and grid spacing ($\sim$30\,m here), and preserves native temporal cadence for sub-daily fire and weather observations.

\section{System Architecture}

FireDataForge provides an event-centered workflow (Fig.~\ref{fig:pipeline}). Users provide a target \ac{crs}, spatial resolution, and an \ac{mtbs} Event ID\footnote{An \ac{mtbs} Event ID can be looked up by fire name, year, and state through the MTBS Data Explorer at \url{https://www.mtbs.gov/viewer/}.}; the framework resolves the fire duration and domain, derives a default grid extent and temporal window, and allows both to be customized. Querying by exact Event ID makes event matching unambiguous and reproducible because one ID resolves to one fire.

The framework retrieves data through category-specific processing components for fire behavior, weather, vegetation and land surface characteristics, terrain, built environment, and satellite imagery. New sources such as MODIS products or ERA5-Land reanalysis can be integrated without altering the overall pipeline. Retrieval tasks are parallelized when possible, while rate-limited sources are processed sequentially.

In the final stage, all datasets are harmonized spatially and temporally. Spatially, each source is reprojected, resampled by data type (Table~\ref{tab:sources}), and rasterized when needed; all layers use the target grid except \ac{hrrr}, which is kept on a $\sim$500\,m output grid with the same extent and projection because upsampling would add no detail. Temporally, \ac{feds}/\ac{viirs} fire observations remain 12-hourly, \ac{hrrr} weather hourly, and static layers single-frame. At consumption time, a loader advances each source's datetime cursor to the latest valid frame whose timestamp does not exceed the simulation clock, otherwise holding the previous frame. Independent cursors let hourly weather update between 12-hour fire snapshots while keeping static layers fixed and skipping missing observations without interpolation, temporal resampling, or future-looking access. Output metadata records \ac{crs}, resolution, units, nodata values, per-frame timestamps, and source attribution; the full schema is documented in the released repository.

\section{Data Sources and Processing}

Table~\ref{tab:sources} summarizes the integrated sources, native spatial and temporal resolutions, output layers, and resampling methods; below we describe only processing details not captured there.

\begin{table*}[!t]
\caption{Integrated data sources and harmonization methods.}\label{tab:sources}
\centering
\small
\setlength{\tabcolsep}{4pt}
\renewcommand{\arraystretch}{1.0}
\begin{tabular}{>{\raggedright\arraybackslash}p{2.5cm}>{\raggedright\arraybackslash}p{5.4cm}ll>{\raggedright\arraybackslash}p{3.8cm}>{\raggedright\arraybackslash}p{2.4cm}}
\toprule
\textbf{Dataset} & \textbf{Description} & \textbf{Spatial} & \textbf{Cadence} & \textbf{Output layer(s)} & \textbf{Resampling} \\
\midrule
FEDS~\cite{chenCaliforniaWildfireSpread2022,chenFEDSMTBSMTBSconstrainedFEDS2026} & VIIRS-derived perimeter and fireline snapshots & 375\,m & 12-hourly & \texttt{burn\_perimeter}, \texttt{fireline} & Polygon raster \\
VIIRS Active Fire~\cite{schroederNewVIIRS3752014} & Hotspot Fire Radiative Power (FRP) observations & 375\,m & 12-hourly & \texttt{frp\_daytime}, \texttt{frp\_nighttime} & Gaussian splat \\
3DEP~\cite{USGeologicalSurvey} & Elevation and RGB hillshade & 1\,m & Static & \texttt{elevation}, \texttt{terrain\_rgb} & Bilinear \\
LANDFIRE~\cite{u.s.departmentoftheinteriorgeologicalsurveyLANDFIRE2402025} & Canopy bulk density and cover & 30\,m & Static & \texttt{canopy\_bulk\_\allowbreak density}, \texttt{canopy\_cover} & Bilinear \\
HRRR~\cite{dowellHighResolutionRapidRefresh2022} & 2\,m relative humidity and 10\,m winds & 3\,km & Hourly & \texttt{r2}, \texttt{u10}, \texttt{v10} & Bilinear \\
Global Building Atlas~\cite{zhuGlobalBuildingAtlasOpenGlobal2025} & Building-height grid & 3\,m & Static & \texttt{building\_height} & Area-weighted mean \\
WorldCover~\cite{zanagaESAWorldCover102022} & 11-class global land cover & 10\,m & Static & \texttt{landcover} & Nearest neighbor \\
Sentinel-2 LAI~\cite{mukherjeeBackgroundClimateSocioeconomic2026} & Leaf Area Index (LAI) & 10\,m & Static & \texttt{lai} & Bilinear \\
Sentinel-2 Cloudless Mosaic~\cite{druschSentinel2ESAsOptical2012} & Annual cloud-free RGB mosaic & 10\,m & Annual & \texttt{sentinel2\_rgb} & Nearest neighbor \\
Global WUI~\cite{schugGlobalWildlandUrban2023} & Building-vegetation WUI classes & 10\,m & Static & \texttt{wui} & Nearest neighbor \\
NIFC IFPH & Prior-5-year burn polygons & Vector & Static & \texttt{recent\_burn} & Polygon raster \\
\bottomrule
\end{tabular}
\end{table*}

\textbf{Fire behavior (\acs{feds}).} FireDataForge identifies the active fire-progression window from changes between consecutive perimeters and filters timesteps with no perimeter growth, which can occur when no expansion is detected between satellite overpasses. For each active fireline segment, it computes \texttt{fireline\_max\_frp} as the maximum \ac{viirs} \ac{frp} within a 360\,m buffer (12 pixels on the 30\,m grid), linking fire-front geometry with observed radiative intensity.

\textbf{Active fire hotspots (\acs{viirs} via \acs{nasa} \acs{firms}).} \ac{viirs} active-fire detections from \ac{nasa} \ac{firms}~\cite{daviesFireInformationResource2009} provide point observations of radiative power. FireDataForge separates detections into \texttt{frp\_daytime} and \texttt{frp\_nighttime} channels according to daytime and nighttime overpasses and maps each detection to the target grid with a sensor-footprint Gaussian splat normalized to conserve total \ac{frp}, so summing the rasterized footprint recovers the observed value. The footprint matches the \ac{viirs} 375\,m pixel using $\sigma$ equal to half the source-to-target pixel ratio ($6.25$\,px, or $\approx441$\,m full width at half maximum on the 30\,m grid) and is truncated at $\pm3\sigma$.

\textbf{Terrain (\acs{threedep}).} In addition to the bilinearly resampled elevation, a colored hillshade \texttt{terrain\_rgb} is computed from the elevation using standard cartographic illumination (azimuth 315\textdegree, elevation 45\textdegree).

The remaining sources are processed per Table~\ref{tab:sources} without source-specific post-processing.

\textbf{Resampling and target resolution.} Resampling is data-type specific and applied whether a source is finer or coarser than the target: nearest-neighbor preserves categorical classes without inventing mixed labels, bilinear handles continuous layers, and area-weighted averaging summarizes building height. Resampling never creates information, so overly coarse targets discard fine variation in terrain, fuels, buildings, and \ac{wui}; users should set the target near the finest source their analysis requires. Per-layer \texttt{native\_resolution} metadata keeps this scale mismatch explicit.

\section{Evaluation}

\begin{figure*}[!t]
\centering
\includegraphics[width=\linewidth]{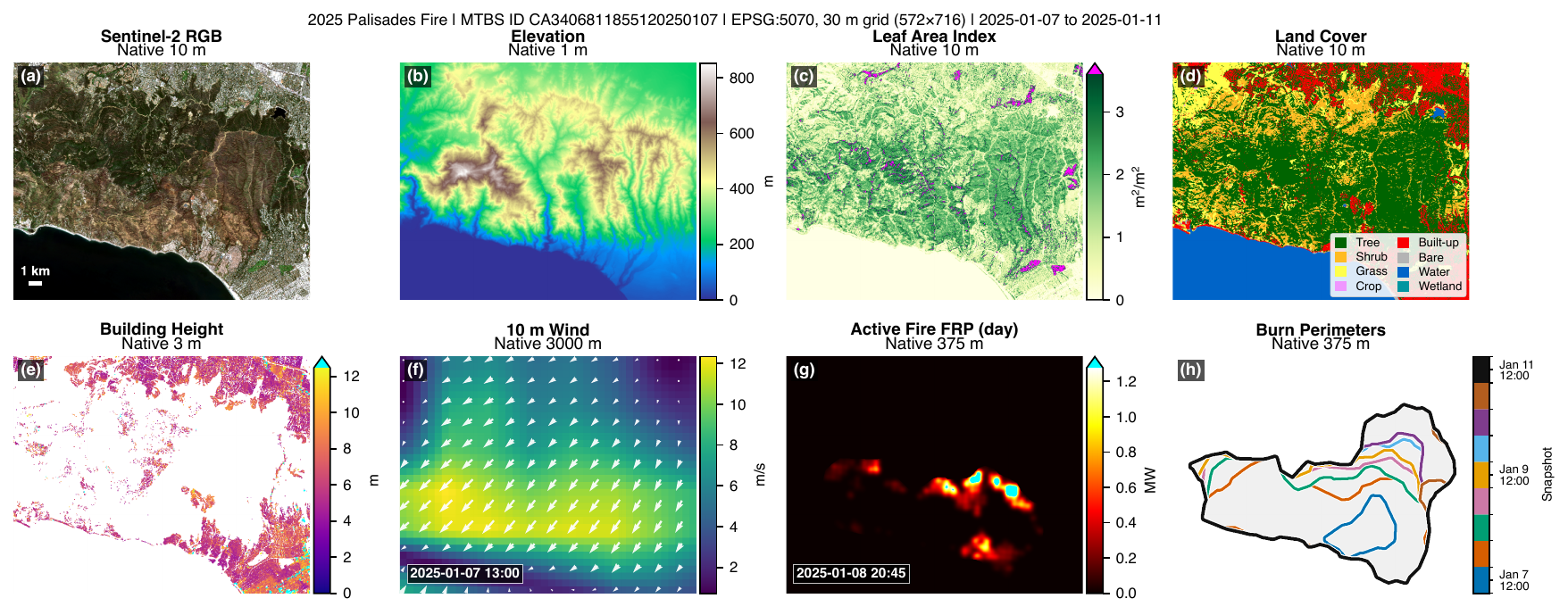}
\caption{Representative multi-source FireDataForge output for the 2025 Palisades Fire~\cite{woolcottAngelesCountyFlames2025}: context imagery (a), terrain (b), vegetation and fuels (c, d), built environment (e), weather (f), and fire behavior (g, h). Panel subtitles report native source resolution; all layers share the event extent and projection, with \acs{hrrr} weather retained on its coarser grid.}\label{fig:output_visual}
\end{figure*}

{\sloppy
We validated FireDataForge on 8 historical fire events from 2013--2025: 5 in California (\acs{mtbs} IDs CA3432611848120191010, CA3859812261820171009, CA3982012144020181108, CA3419211810520250108, and the 2025 Palisades Fire CA3406811855120250107) and 3 in Colorado (CO3749610529120180627, CO3901210474920130611, CO4020310623920201014). The events span chaparral, conifer forest, grassland, varied wind and ignition conditions, built-environment contexts, durations, and spatial scales, testing integration across heterogeneous fire contexts rather than national climatology. Per-event metadata---fire name, year, state, \ac{mtbs} ID, grid size, and temporal window---is released for exact reproduction. We use the Palisades Fire to illustrate harmonized multi-layer outputs (Fig.~\ref{fig:output_visual}).\par}

\subsection{Data Quality and Alignment}

For evaluation, datasets were aligned to EPSG:5070 (NAD83 / CONUS Albers Equal Area) at 30\,m, with \ac{hrrr} retained on its coarser grid. Fire perimeters, \ac{frp} fields, and weather variables showed consistent spatial and temporal alignment across native projections including Lambert Conformal Conic, WGS84, and \ac{utm}.

End-to-end runtime was dominated by network I/O. Each event was processed single-threaded (\texttt{--workers 1}) inside an 8-core, 32\,GB SLURM allocation on a node with dual Intel Xeon Gold 6248 CPUs @ 2.50\,GHz. Grids of $361\times388$ to $1397\times1448$\,px (0.14--2.0\,M pixels) and 5--24 perimeter timesteps processed in $13.5$--$269.1$\,s with a cold cache (median $\approx125$\,s) and $12.7$--$221.7$\,s with a warm cache (\ac{hrrr} GRIB2, \ac{wui} tiles, \ac{firms} slices, and the \ac{mtbs} fire list from local disk, with \ac{gee} and NIFC still accessed over the network; median $\approx86$\,s). Cold runs were driven mainly by 49--294 hourly \ac{hrrr} GRIB2 downloads from \ac{noaa} \ac{aws} S3, so wall time tracks upstream latency more than grid size. All 48 timed runs completed with zero failed layers; all events produced 20 layers except the 2013 fire, which predates \ac{hrrr} and Sentinel-2 and produced 15. We report per-event medians over three repetitions, as upstream latency occasionally inflates a single run; per-event timings are released in the reproducibility bundle.

\subsection{Quantitative Validation}

We quantified harmonization fidelity against independently constructed native-resolution references aggregated to the target grid by data type (majority vote for categorical layers, mean for continuous). Across the 8 events, coordinate round-trip error was negligible ($3.0\times10^{-9}$\,m), indicating internally consistent \ac{crs} and axis-order handling; \ac{frp} was conserved within $0.004\%$ after Gaussian splatting; elevation RMSE was $4.27$\,m against mean-aggregated \acs{threedep}; and land-cover/\ac{wui} accuracies were $0.944$/$0.998$ respectively against majority-aggregated references. These tests do not measure independent cross-source registration or upstream-product accuracy, but they show that reprojection, rasterization, and resampling preserve key source characteristics. Per-event values are provided in the released bundle.

\section{Conclusion}

FireDataForge automates retrieval and spatiotemporal harmonization of wildfire-related geospatial data spanning satellite observations, terrain, vegetation, weather, and the built environment. Its analysis-ready outputs lower preprocessing barriers for fire behavior research, educational visualization, machine learning, and \ac{ai}-assisted wildfire analysis while preserving source attribution and native-resolution context.

Because outputs include arrays plus timestamps, units, nodata values, \ac{crs}, and source attribution, users can reuse layers without source-specific retrieval code. Its modular architecture supports new data sources, and batch processing demonstrates scalability. Future work will add web-based visualization, integrate with fire-spread models like PyTorchFire~\cite{xiaPyTorchFireGPUacceleratedWildfire2025}, and incorporate human-system dimensions such as evacuation dynamics.

\section*{Code and Artifact Availability}
{\interlinepenalty=10000
Source code is available at \url{https://github.com/xiazeyu/FireDataForge} and archived on Zenodo (\url{https://doi.org/10.5281/zenodo.20744740}). The reproducibility bundle is archived at \url{https://doi.org/10.5281/zenodo.20743743}.\par}

\section*{Acknowledgment}
This work is supported by the University of Virginia's Environmental Institute. PNNL is operated for DOE by Battelle Memorial Institute under contract DE-AC05-76RL01830. This product contains modified Copernicus Sentinel data (2017--2025), used to derive the Sentinel-2 surface-reflectance imagery and \ac{lai} layers.

\balance
\bibliographystyle{IEEEtran}
\bibliography{references}

\end{document}